\documentclass[11pt]{JHEP3}
\usepackage{amsmath,amssymb,graphicx,epsfig}
\usepackage{psfrag}

\newcommand{\newc}{\newcommand}
\newc{\beq}{\begin{equation}}
\newc{\eeq}{\end{equation}}
\newc{\beqa}{\begin{eqnarray}}
\newc{\eeqa}{\end{eqnarray}}
\newc{\bc}{\begin{center}}
\newc{\ec}{\end{center}}
\newc{\nonr}{\nonumber}
\newc{\PD}{\partial}
\newc{\bd}{\begin{description}}
\newc{\ed}{\end{description}}
\newc{\benu}{\begin{enumerate}}
\newc{\eenu}{\end{enumerate}}
\newc{\bi}{\begin{itemize}}
\newc{\ei}{\end{itemize}}
\newc{\ra}{\rightarrow}
\newc{\bis}{\begin{itemstep}}
\newc{\eis}{\end{itemstep}}
\newc{\lag}{\cal{L}}
\newc{\LH}{\hat{L}}
\newc{\RH}{\hat{R}}
\newc{\SL}{\not\!}
\newc{\mtr}{\mathrm {tr}}
\newc{\ol}{\overline}
\newc{\Gf}{\frac{G_F}{\sqrt 2}}
\newc{\nonu}{0\nu\beta\beta}
\newc{\half}{\frac{1}{2}}
\newc{\Mred}{M_{--}/100\,\mathrm{GeV}}
\newc{\ugl}{\ol{u_L}^i\gamma^{\mu}u_{L}^i}
\newc{\ugr}{\ol{u_R}^i\gamma^{\mu}u_{R}^i}
\newc{\egl}{\ol{e}^k\gamma_{\mu}\LH e^k}
\newc{\egr}{\ol{e}^k\gamma_{\mu}\RH e^k}
\newc{\dgl}{\ol{d}^i\gamma^{\mu}\LHd^i}
\newc{\dgr}{\ol{d}^i\gamma^{\mu}\RHd^i}
\newc{\alo}{\alpha_1}
\newc{\alt}{\alpha_2}
\newc{\alth}{\alpha_3}

\title{Testing Radiative Neutrino Mass Generation at the LHC}
\author{Chian-Shu Chen, Chao-Qiang Geng \\
Department of Physics, National Tsing Hua University, Hsinchu, Taiwan 300}
\author{John N. Ng, Jackson M. S. Wu \\ 
Theory Group, TRIUMF, 4004 Wesbrook Mall, Vancouver, B.C. Canada V6T 2A3}

\abstract{We investigate in detail a model that contains an additional $SU(2)$
singlet and triplet scalar fields than the Standard Model (SM). This allows the 
radiative generation of Majorana neutrino masses at two-loop order with the help of 
doubly charged Higgs bosons that arise from the extended Higgs sector. The 
phenomenology of the Higgs and neutrino sectors of the model is studied. We give
the analytical form of the masses of scalar and pseudoscalar bosons and their 
mixings, and the structure of the active neutrino mass matrix. It is found that the 
model accommodates only normal neutrino mass hierarchy, and that there is a large 
parameter space where the doubly charged Higgs can be observed at the Large Hadron 
Collider (LHC), thereby making it testable at the LHC. Furthermore, the 
neutrino-less double beta ($\nonu$) decays arise predominantly from exchange 
processes involving the doubly charged Higgs, whose existence is thus unmistakable 
if $\nonu$ decays are observed. The production and decays of the doubly charged 
Higgs are analyzed, and distinct and distinguishing signals are discussed.} 


\preprint

\begin{document}

\section{Introduction}

The origin of small active neutrino masses remains one of the most challenging 
problem in physics. The small neutrino masses generated through the seesaw 
mechanism is popularly viewed as heralding new physics at scales larger than 
$10^{12}$~GeV, and thus provide a window to Grand Unified Theories (GUTs) with 
or without supersymmetry. Crucial to the construction is the introduction of 
heavy Standard Model (SM) singlet fermions commonly known as sterile neutrinos.
 
Recently, the idea of extra spatial dimensions together with brane world scenarios 
offers a very different perspective to the question of neutrino masses. Here their 
smallness results from either the suppression factors associated with the 
relatively large extra dimensions, or from the small overlap between the 
wave functions of the sterile neutrinos in the extra dimensions. 

It is interesting to note that these different perspectives can be incorporated 
into a single framework in brane world scenarios; a recent discussion can be found 
in~\cite{HS}. However the existence of sterile neutrinos is required in both 
constructions which, along with the value of their masses, are all important 
questions by themselves. To date the best information on light sterile neutrinos 
comes from cosmological considerations; direct experimental tests are very 
challenging due to the fact that they have no SM interactions.

It is well known that the masses of active neutrinos can be generated without 
sterile right-handed (RH) neutrinos via quantum loop effects. Without the RH 
states there are no Dirac couplings of the SM lepton doublet to the Higgs fields, 
and consequently the active neutrinos can only have Majorana masses. The prototype 
model was constructed in~\cite{Zee} where there is an extended Higgs sector, and 
the gauge symmetry is that of the SM. Crucial to the construction was the use of an 
$SU(2)$ singlet Higgs field with a nontrivial hypercharge. Unfortunately the model 
gives rise to bimaximal neutrino mixings which is disfavored by the most recent 
neutrino data (for a recent review see~\cite{nurev}). More realistic neutrino 
masses can be obtained using doubly charged Higgs fields~\cite{Zee2}. 

In our construction, we keep the SM gauge group and extend the Higgs sector by 
adding both an $SU(2)$ triplet and a doubly charged singlet field. We also 
postulate that lepton number violating effects take place only in the scalar 
potential, while the rest of the Lagrangian respect lepton number. A brief 
discussion of our model has already appeared in~\cite{P1} where we showed how 
naturally small neutrino masses can arise from just two-loop radiative corrections. 
In this paper we will give a detail discussion of rich scalar phenomenology of the 
model. In particular the signals at the Large Hadron Collider (LHC) are investigated.

The rest of the paper is organized as follow. In Sec.~2 we describe in detail our 
model. We work out the constraints on the vacuum expectation value and other
parameters that control the mass of the two physical doubly charged Higgs bosons 
$P_{1,2}^{\pm\pm}$. We show that at least one of the doubly charged Higgs can have 
mass at the electroweak scale if we demand the theory be perturbative up to the TeV 
scale. 
In Sec.~3 we discuss in detail the phenomenology of the neutrino sector in our 
model. We examine closely the neutrino mass matrix and the constraints from the 
oscillation data. We show that normal hierarchy arises naturally in our model, and 
we place constraints on the neutrino-lepton Yukawa couplings. Lastly, we discuss 
the implications these constraints have on the $\nonu$ decays of nuclei in our 
model. In Sec.~4 we discuss the phenomenology of the doubly charged Higgs 
production at the LHC, and their decays. We show that the decay pattern of the doubly
charged Higgs in our model can be very different, and can therefore be used to  
distinguish our model from others that also contain doubly charged Higgs. Sec.~5 
contains our conclusions.

\section{A minimal model with radiative neutrino mass generation.}

The model is based on the SM gauge symmetry with an extended Higgs sector and 
minimal matter content. Group theory dictates that only $SU(2)$ singlets and 
triplets are allowed for the generation of Majorana masses for the neutrinos. 
Besides the SM Higgs doublet given by
\begin{equation}
\phi = 
\begin{pmatrix}
\phi^0 \\
\phi^-
\end{pmatrix}_{-1} \,, \qquad
\widetilde{\phi} = i\tau_2 \phi \,,
\end{equation}
we introduce a complex triplet Higgs $T$
represented by a $2\times 2$ matrix
\begin{equation}
T = 
\begin{pmatrix}
T^0 & \frac{T^-}{\sqrt{2}} \\
\frac{T^-}{\sqrt{2}} & T^{--}
\end{pmatrix}_{-2} \,,
\end{equation}
as well as a complex singlet scalar $\Psi^{++}_4$. The subscripts denote the weak 
hypercharges of the fields as given by the relation $Q=T_3+\half Y$. The most 
general potential for the scalar fields is given by
\begin{align}\label{HLVV}
V(\phi,T,\psi) &= 
-\mu^2\phi^{\dag}\phi + \lambda_{\phi}(\phi^{\dag}\phi)^2
-\mu^2_TTr{(T^{\dag}T)} + \lambda_T[Tr(T^{\dag}T)]^2 
+\lambda'_TTr(T^{\dag}TT^{\dag}T) \notag \\
&\quad 
+m^2\Psi^{\dag}\Psi+ \lambda_{\Psi}(\Psi^{\dag}\Psi)^2 
+\kappa_1Tr(\phi^{\dag}\phi T^{\dag}T) + \kappa_{2}\phi^{\dag}TT^{\dag}\phi
+\kappa_{\Psi}\phi^{\dag}\phi\Psi^{\dag}\Psi \notag \\
&\quad
+\rho\,Tr(T^{\dag}T\Psi^{\dag}\Psi) 
+\left[\lambda(\widetilde{\phi}^{T}T\widetilde{\phi}\,\Psi)
-M(\phi^{T}T^{\dag}\phi)+h.c.\right]\,.
\end{align}
One can assign $\Psi$ a lepton number 2, and $T$ a lepton number 0. Then terms in 
the square brackets at the end of Eq.~\eqref{HLVV} contain lepton number violating 
interactions. We will take both $\mu^2$, $\mu_{T}^2$ to be positive so that 
spontaneous symmetry breaking (SSB) takes place. Minimizing the potential gives us 
the VEVs: $\langle \phi^0 \rangle \equiv \frac{v}{\sqrt 2}$ and 
$\langle T^0 \rangle \equiv \frac{v_T}{\sqrt 2}$.

With the additional fields $T$ and $\Psi$, two new Yukawa terms can be constructed 
that are allowed by the gauge symmetry. The first is 
$Y_{ab}\overline{l^c_{aR}}l_{bR}\Psi$, which is lepton number conserving. Here 
$a,b$ are family indices and $l_a$ is a lepton singlet. The second is $LLT$ that 
violates lepton number, which we assume not to occur at the tree 
level~\footnote{There are several ways to naturally suppress the Yukawa couplings 
of $L_{a}L_{b}T$. One way is to embed the model in a 5-dimensional set-up and 
compactify the extra dimensions on an orbifold $S_1/Z_2$. The lepton and Higgs 
fields are then assigned with different orbifold parities to forbid the $LLT$ term 
while still allows the $ll\psi$ term. Another way is to further extend the Higgs 
sector by including a second Higgs doublet and then employ an appropriate discrete 
symmetry. The $LLT$ term will be generated radiatively after symmetry breaking, but
is small.}. The absence of this term frees one from having to put in by hand a very
small value of $v_T$ in the eV range that plagues other Higgs models of neutrino 
masses. Adding in the SM terms and the covariant derivatives of $T$ and $\Psi$ we 
have a complete renormalizable model.
 
 From Eq.~(\ref{HLVV}) it can be seen that the various Higgs fields will mix among
themselves. In particular, the pair $\mathfrak{Re}\,\phi^0$ and 
${\mathfrak{Re}}\,T^0$ will mix that give rise to two physical neutral scalars, 
$h^0$ and $P^0$, and the pair $\mathfrak{Im}\,\phi^0$ and $\mathfrak{Im}\,T^0$ will 
mix with one combination that is eaten by the $Z$ boson to leave a physical 
pseudoscalar $T^0_a$. Similarly for the charged states $\phi^{\pm}$ and $T^{\pm}$, 
one combination will be eaten by the $W$ bosons leaving only a pair of singly 
charged $P^{\pm}$ scalars. Finally the weak eigenstates $T^{\pm\pm}$ and 
$\Psi^{\pm\pm}$ will also mix to form physical states $P_1^{\pm\pm}$ and 
$P_2^{\pm\pm}$, with the mixing angle denoted by $\delta$, and masses $M_1$ and 
$M_2$ respectively. All the masses and mixing angles are free parameters in our 
model, which we can use to replace some of the parameters in $V(\phi,T,\Psi)$. They 
are to be determined experimentally. In summary, the physical spin~0 particles 
consist of a pair of singly charged bosons, $P^\pm$, two pairs of doubly charged 
bosons $P_1^{\pm\pm}$ and $P_2^{\pm\pm}$, a pair of Higgs scalars $h^0$ and $P^0$, 
and a pseudoscalar $A^0$. 

The value of $v_T$ is constrained by the electroweak phenomenology. After the
electroweak symmetry breaking, the $W$ and $Z$ bosons pick up masses at the tree 
level given by 
\begin{equation}
M_W^2 = \frac{g^2}{4}(v^2+2v_T^2) \,, \qquad
M_Z^2 = \frac{g^2}{4\cos^2\theta_{W}}(v^2+4v_T^2) \,,
\end{equation}
where we have used standard notations, and the tree level relation 
$e=g\sin\theta_W$ holds. From the Particle Data Group (PDG) we have 
$\rho=1.002^{+.0007}_{-.0009}$~\cite{PDG} and  $M_W=80.41$~GeV~\cite{CDF}. This 
implies that $v_T<4.41$~GeV. We will see below that this is a controlling scale for 
the neutrino masses.

In the limit where $\lambda$, $M \ra 0$ the model conserves lepton number, and is 
thus technically natural. Since $\lambda$ is a dimensionless coupling, it is 
expected to be of order unity ($< 4\pi$) so that perturbation is valid, which is 
assumed throughout this paper. The value of $M$ is important in setting the scale 
of lepton number violation. It enters in the conditions for minimizing the scalar
potential $V(\phi,T,\Psi)$:
\begin{align}\label{Vmin}
-&\mu^2 + \lambda_{\phi}v^2 + \half\kappa_{+}v_T^2 - \sqrt{2}M v_T = 0 \,, \\
-&\mu^2_T + \lambda_{+}v^2_T + \half\kappa_{+} v^2
  -\frac{v^2}{\sqrt{2}}\left(\frac{M}{v_T}\right) = 0 \,,
\end{align}
where $\kappa_{+}=\kappa_1+\kappa_2$ and $\lambda_{+}=\lambda_T+\lambda_T^{\prime}$.
Taking $v$ to be of the electroweak scale and $v_T/v \approx 0.02$, the interesting 
limiting cases are:
\benu
\item $M \sim v_T$: The minimum conditions, Eq.~(\ref{Vmin}), can be naturally 
satisfied without fine tuning between the parameters if $\mu_T \sim v_T$.

\item $M \sim v \gg v_T$: To satisfy the minimum conditions, $\mu_T^2 \sim v^3/v_T$
is required. This appears to be unnatural although not forbidden.

\item $M > v$: The minimum conditions can only be satisfied by tuning the 
dimensionful and/or the dimensionless couplings. We will not consider this case.
\eenu
For convenience define $\omega\equiv\frac{M}{\sqrt{2}v_T}$. Then Case A and B 
correspond to $\omega \sim 1$ and $\omega \gg 1$ respectively. Qualitatively, we see
that $\omega \lesssim 1$ is more natural in our model. We will therefore concentrate
mostly in this region of the parameter space below.

We now turn to the masses of the physical scalar and pseudoscalar particles in our
model. The mass of the singly charged Higgs boson is given by
\begin{equation}
\label{MassP}
M_{P^\pm}^2 = \left(\omega-\frac{\kappa_2}{4}\right)(v^2+2v_T^2) \,.
\end{equation}
For $\omega \sim 1$, we expect the charged Higgs to have mass in the 100 to 1000 
GeV range.

For the two doubly charged scalars, their masses are given by 
\begin{equation}\label{DCH}
M_{P_{1,2}}^2 = \half\left[a + c \mp\sqrt{4b^2 + (c-a)^2}\right] \,, 
\end{equation}
where the $P_1$ state takes the upper sign, and
\begin{equation}\label{MMabc}
a = \half(2\omega - \kappa_2)v^2 - \lambda_T' v_T^2 \,, \qquad
b = \half \lambda v^2 \,, \qquad
c = m^2 + \half(\kappa_\Psi v^2 + \rho v_T^2) \,.
\end{equation}
Note that $m$ ($m^2 > 0$) is a mass parameter for the singlet and should not be 
confused with the physical mass. As such it is in general not constrained. 

Consider now Case A. In the limit where $m$ is large ($m \gg v$), we have from 
Eqs.~\eqref{DCH} and~\eqref{MMabc}:
\begin{align}
M_{P_1}^2 &= 
\half(2\omega - \kappa_2)v^2 - \lambda_T' v_T^2 
+ \mathcal{O}\left(\frac{v^4}{m^2}\right) \,, \\
M_{P_2}^2 &= 
m^2 + \half(\kappa_\Psi v^2 + \rho v_T^2)
+ \mathcal{O}\left(\frac{v^4}{m^2}\right) \,.
\end{align}
We see in this limit, $M_{P_1}$ saturates to an m-independent value,
$\sqrt{\half(2\omega-\kappa_2)v^2-\lambda_T' v_T^2}$, which is also its maximal
value for a given set of model parameters. On the other hand, $M_{P_2}$ increases 
as $m$, which means the $P_2$ state will be too heavy to be of interest to the LHC 
in the large $m$ limit. 

\EPSFIGURE{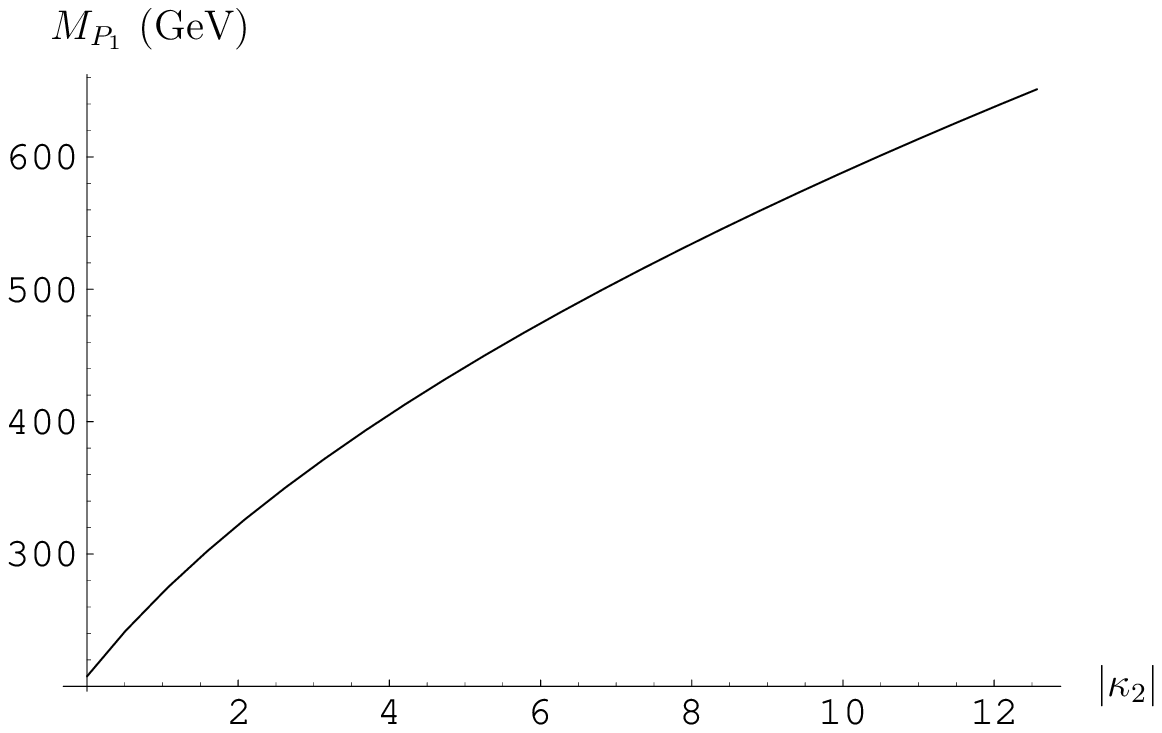,width=3.6in}
{Maximum value of $M_{P_1}$ for $v_T = M = 4$~GeV, and $|\lambda_T'|$ set to $4\pi$.
\label{fig:P1max}}

\EPSFIGURE{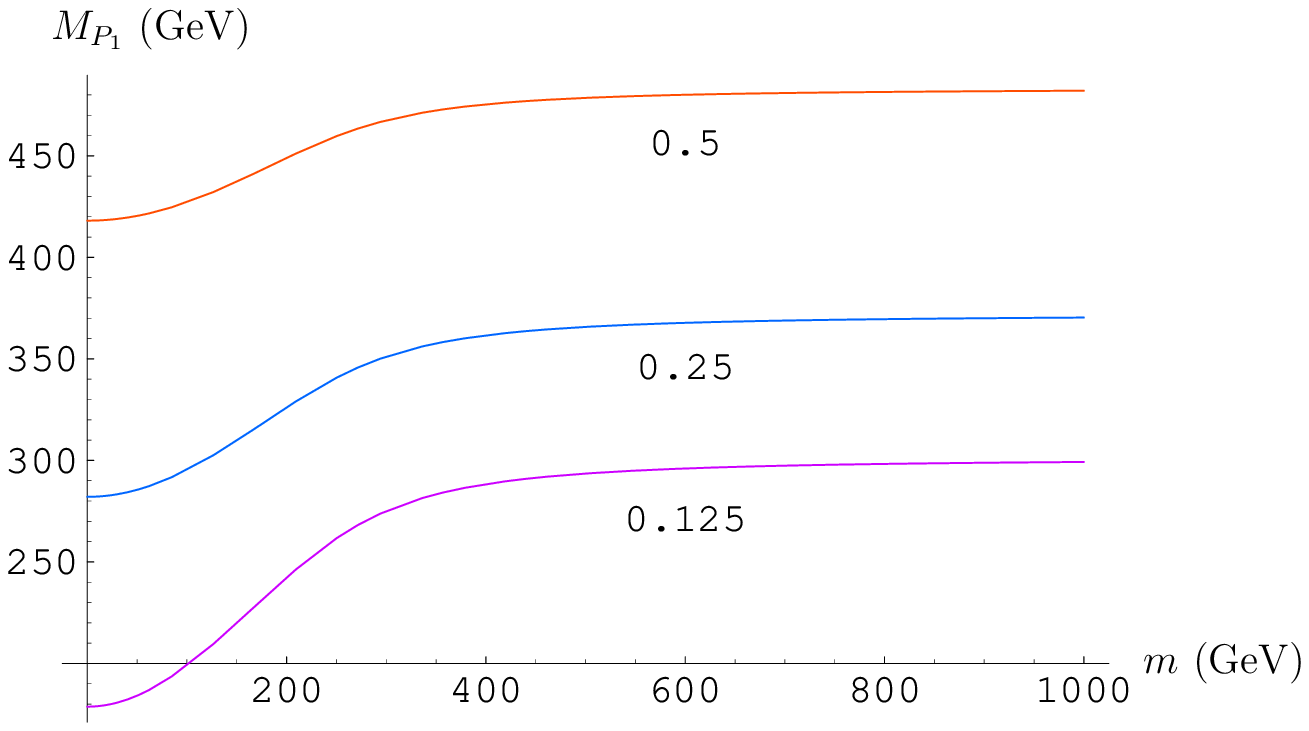,width=3.6in}
{$M_{P_1}$ as a function of $m$ for $|\kappa_2| = 0.5,\,0.25,\,0.125$ in units of 
$4\pi$, with $v_T = M = 4$~GeV and $\lambda = -\lambda_T' = 1$.
\label{fig:P1vsm}}

In Fig.~\ref{fig:P1max} we plot the maximal value of $M_{P_1}$ as a function of 
$|\kappa_2| = -\kappa_2$, with $v_T$ and $M$ set to 4~GeV. The coupling $\lambda_T'$
is set to $|\lambda_T'| = -\lambda_T' = 4\pi$, the upper limit under which 
perturbation is expected to be valid. In Fig.~\ref{fig:P1vsm} we plot $M_{P_1}$ as a
function of $m$ for three different values of $|\kappa_2|$, with all dimensionless 
couplings kept perturbative. We see clearly here the saturation of $M_{P_2}$ at 
large values of $m$. Figs.~\ref{fig:P1max} and~\ref{fig:P1vsm} show that the mass 
range of the $P_1$ state is well within the reach of the LHC, its existence is thus 
a testable feature of our model at the LHC. 

Note that if we take the opposite limit where $m \ra 0$, two weak scale doubly
charged scalars are possible. For example, if we take $\omega=1$, $\lambda=1$, 
$\kappa_{\Psi} = -\kappa_2 = 2$, and $\rho = 2\lambda_T' = 2$, we get 
$M_{P_1} = 219.3$~GeV and $M_{P_2} = 365.8$~GeV. However, this will not hold in
Case B. In this case the $P_2$ state will be heavy with mass above a TeV, while 
the $P_1$ state remains at the weak scale.

The doubly charged scalars form a two-level system in which the mass and weak 
eigenstates are related by
\begin{equation}\label{DCM}
\begin{pmatrix}
P_1^{\pm\pm} \\
P_2^{\pm\pm} 
\end{pmatrix}
=
\begin{pmatrix}
\cos\delta & \sin\delta \\
-\sin\delta & \cos\delta 
\end{pmatrix}
\begin{pmatrix}
T^{\pm\pm} \\
\Psi^{\pm\pm}
\end{pmatrix} \,.
\end{equation}
The mixing angle $\delta$ is a measurable physical parameter. It is given by
\begin{align}\label{dcmixing}
\sin 2\delta 
&= \left[1 + \frac{(c-a)^2}{4b^2}\right]^{-\half} \notag \\
&= \left[1 + \left(\frac{2m^2 + (2\lambda_T' + \rho)v_T^2}{2\lambda v^2} 
                  +\frac{\kappa_2 + \kappa_\Psi}{2\lambda}
                  -\frac{\omega}{\lambda}\right)^2\right]^{-\half} \,,
\end{align}
where $a$, $b$ and $c$ are those given in Eq.~\eqref{MMabc}. For Case A, if 
$m^2 \lesssim v^2$, the mixing can be large and close to maximal. But if 
$m^2 \gg v^2$, the mixing will be small, which is expected since the two states 
are widely split. For case B, large mixing can be achieved only if a cancellation 
occurs between the various parameter in Eq.~(\ref{dcmixing}).

We now turn to the three physical neutral scalar bosons in our model. The 
pseudoscalar $T^0_a$ has mass given by
\begin{equation}\label{psmass}
M_{T^0_a}^2 = \half\omega(v^2+ 4v_T^2) \,.
\end{equation}
Note that if $\omega \ra 0$, $T^0_a$ becomes a Majoron. 

The masses of the neutral scalars $h^0$ and $P^0$ again have the general form
\begin{equation}\label{smass}
M_{h^0,P^0}^2 = \half\left[a' + c' \mp\sqrt{4b^{\prime 2} + (c'-a')^2}\right] \,,
\end{equation}
where the $h^0$ state takes the upper sign, and
\begin{equation}\label{sMMabc}
a' = \lambda_\phi v^2 \,, \qquad
b' = \left(\half\kappa_{+}-\omega\right)v v_T \,, \qquad
c' = \lambda_{+} v_T^2 + \half\omega v^2 \,.
\end{equation}
The physical neutral scalars also form a two-level system in which the mass and 
weak eigenstates mix, with the mixing angle, $\vartheta$ given by
\begin{align}
\sin{2\vartheta}
&= \left[1 + \frac{(c'-a')^2}{4b^{'2}}\right]^{-\half} \notag \\
&= \left[1 + \frac{v^2}{16v_T^2}\left(
             \frac{2\lambda_+}{\kappa_{+}-2\omega}\frac{v_T^2}{v^2}
            +\frac{\omega-2\lambda_\phi}{\kappa_{+}-2\omega}\right)^2
   \right]^{-\half} \,.
\end{align}
It can be seen that $\vartheta$ is of order $v_T/v$ for both Case A and B. 


For Case A, $T_a^0$ and $P^0$ can both be light and almost degenerate.
If they are lighter than half the Z boson mass they
will contribute to its invisible width~\cite{GN}:
\begin{equation}
\Gamma(Z^0\ra P^0\,T^0_a) = 
\frac{G_FM_Z^3}{6\sqrt{2}\pi}\left( 1-2\frac{M_{p^0}^2+M_a^2}{M_Z^2}\right)^3 \,,
\end{equation}
where the notation is standard. Demanding that this contributes less than 150 MeV 
to the invisible width we obtain $|\omega| > 0.016$. For Case B, all the neutral 
bosons have weak scale masses and the above limit does not apply. However, we still 
expect $P^0$ and $T^0_a$ to be close in mass.
 
We summarize our findings on the masses of the scalar and pseudoscalar bosons in 
our model:
\benu
\item $M \sim v_T$: The mass of $P_{1,2}^{\pm\pm}$ is expected to be greater than
$200$~GeV if $m \lesssim 1$~TeV. But if $m$ is much larger than that, the mass of 
$P_1^{\pm\pm}$ will saturate to a constant value which is at most 
$\mathcal{O}(600)$~GeV; the $P_2^{\pm\pm}$ states are expected to be very heavy in 
the large $m$ limit. The singly charged Higgs has a mass at the weak scale that is 
$m$-independent. The masses of neutral Higgs-like bosons are also of the weak scale.
Being the would-be Majoron, $T^0_a$ provides a bound on $\omega$: 
$|\omega| > 0.016$.

\item $M \sim v \gg v_T$: Here, only the mass of $P_1^{\pm\pm}$ is expected to be 
at the weak scale. All the other scalars with the exception of $h^0$ (which is 
mostly a SM Higgs boson) will be too heavy to be of interest at the LHC, since 
their masses are controlled by $\omega$.
\eenu

\section{Neutrino phenomenology and constraints}
\subsection{Two-loop neutrino masses and neutrino oscillations}
A feature of our model is that neutrino masses are generated at the two-loop level, 
and the crucial couplings are the Yukawa terms. It is well known that the Yukawa 
couplings of $\phi$ to fermions are diagonalized by a biunitary transformation
effected via $U_L$ and $U_R$ such that the charged leptons are mass eigenstates. 
Clearly, applying this transformation does not in general diagonalize $Y_{ab}$. 
Hence we expect flavor violating couplings $Y'_{ab}$ between families of RH leptons 
and the physical $P^{++}$ states.~\footnote{In the following we assume that the 
charged leptons are in the mass basis, and so $Y^{\prime} = U_R Y U_R$. For 
notational simplicity we will drop the prime henceforth.} Thus in general, the 
decay modes such as $P^{++}\ra\mu^{+}e^{+}$ must occur. The coupling of $P^{\pm}$ 
to fermions, on the other hand, is similar to SM but scaled by a factor $v_T/v$.

\EPSFIGURE{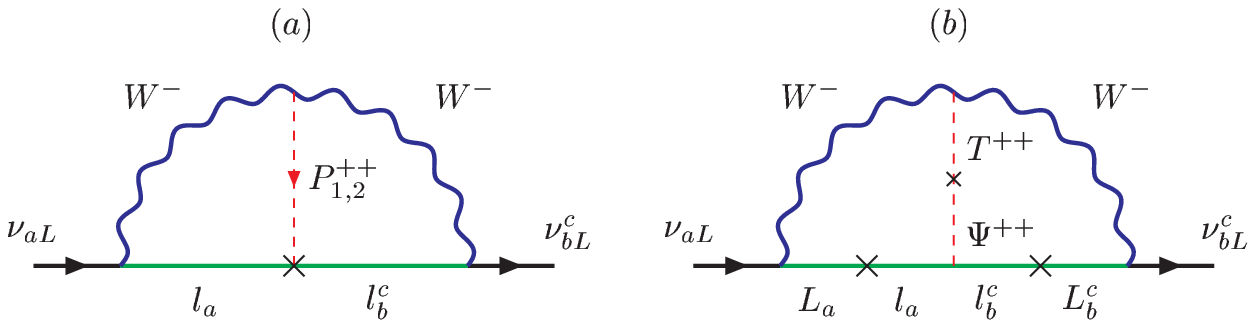,width=4in}
{The two-loop diagram for the neutrino mass in: (a) the mass eigenbasis and 
(b) the weak eigenbasis.
\label{fig:2loop}}

The active neutrino mass matrix can now be calculated. The leading contribution is 
given by the two-loop Feynman diagram depicted in Fig.~\ref{fig:2loop}. After a 
standard but lengthy calculation we find
\begin{equation}
\label{numass}
(m_\nu)_{ab} = \frac{1}{\sqrt{2}}\,g^4 m_a\,m_b\,v_T Y_{ab}\sin(2\delta)  
\left[I(M_W^2,M_{P_1}^2,m_{a},m_{b})- I(M_W^2,M_{P_2}^2,m_{a},m_{b})\right] \,,
\end{equation}
where $a,b=e,\mu,\tau$.
The integral $I$ is given by
\begin{align}
&I(M_{W}^2,M_{P_i}^2,m_a^2,m_b^2) = \notag \\
&\quad
\int\frac{d^4q}{(2\pi)^4}
\int\frac{d^4k}{(2\pi)^4}\frac{1}{k^2-m^2_{a}}\frac{1}{k^2-M_{W}^2} 
    \frac{1}{q^2-M_{W}^2}\frac{1}{q^2-m^2_{b}}\frac{1}{(k-q)^2-M_{P_i}^2} \,.
\end{align}
The integral can be evaluated analytically as in~\cite{2loop}. Note that there is a 
generalized GIM mechanism at work here. This can be seen clearly in the limit 
$M_{P_{1,2}} > M_W$~\cite{2loopl}:
\begin{equation}\label{intg}
I(M_{W}^2,M_{P_i}^2,0,0)\sim
\frac{1}{(4\pi)^4}\frac{1}{M_{P_i}^2}\log^2\!\left(\frac{M_{W}^2}{M_{P_i}^2}\right) 
\,.
\end{equation}
We see that not only is the neutrino mass two-loop suppressed, there is also a 
helicity suppression from the charged lepton masses, whose origin can be clearly 
seen from Fig.~\ref{fig:2loop}(b). It is clear that the internal lepton lines must
have mass insertions since $\Psi^{++}$ only couples to RH leptons. As a result, 
$(m_\nu)_{ee}$ will be vanishingly small. This has important consequences in 
$\nonu$ decays as well as the choice of signatures for the detection of these 
scalars at the LHC.

Explicitly, the neutrino mass matrix is given by
\begin{align}\label{numatrix}
m_\nu &= \widetilde{f}(M_{P_1},M_{P_2}) \times 
\begin{pmatrix} 
m_e^2\,Y_{ee} & m_e m_\mu Y_{e\mu} & m_e m_{\tau}Y_{e\tau} \\
m_e m_\mu Y_{e\mu} & m_{\mu}^2\,Y_{\mu\mu} & m_\tau m_\mu Y_{\mu\tau} \\
m_e m_{\tau}Y_{e\tau} & m_\tau m_{\mu}Y_{\mu\tau} & m_{\tau}^2\,Y_{\tau\tau}
\end{pmatrix} \notag \\
&= f(M_{P_1},M_{P_2}) \times 
\begin{pmatrix}
2.6 \times 10^{-7}\,Y_{ee} & 5.4 \times 10^{-5}\,Y_{e\mu} 
& 9.1\times 10^{-4}\,Y_{e\tau} \\
5.4 \times 10^{-5}\,Y_{e\mu} & 1.1 \times 10^{-2}\,Y_{\mu\mu} 
& 0.19\,Y_{\mu\tau} \\
9.1 \times 10^{-4}\,Y_{e\tau} & 0.19\,Y_{\mu\tau} & 3.17\,Y_{\tau\tau} 
\end{pmatrix} \,,
\end{align}
where
\begin{equation}\label{scale} 
\widetilde{f}(M_{P_1},M_{P_2}) = 
\frac{\sqrt{2}g^4 v_T \sin(2\delta)}{128\pi^4}
\left[\frac{1}{M_{P_1}^2}\log^2\!\left(\frac{M_W}{M_{P_1}}\right)
     -\frac{1}{M_{P_2}^2}\log^2\!\left(\frac{M_W}{M_{P_2}}\right)\right] \,,
\end{equation}
and $f = \widetilde{f}\times(1\mathrm{GeV}^2)$ gives a qualitatively estimate of 
the overall scale of active neutrino masses. Now for normal hierarchy, the neutrino
mass matrix has the following structure
\begin{equation}\label{NH}
\begin{pmatrix} 
\varepsilon^{\prime} & \varepsilon & \varepsilon \\
\varepsilon & 1+\eta & 1+\eta \\
\varepsilon & 1+\eta & 1+\eta 
\end{pmatrix} \,,
\end{equation}
where $\varepsilon$, $\varepsilon'$ and $\eta \ll 1$. Comparing 
Eq.~\eqref{numatrix} to Eq.~\eqref{NH}, we see there is a qualitatively agreement. 

\EPSFIGURE{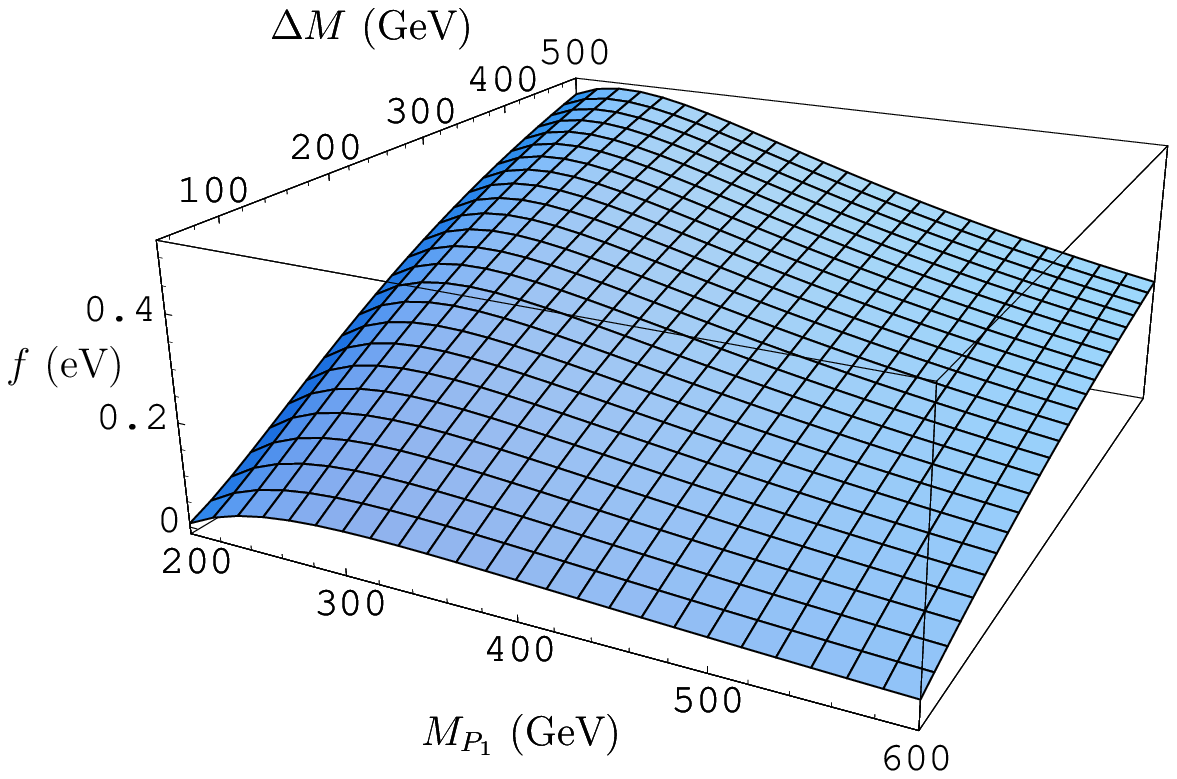,width=3.6in}
{The scale $f(M_{P_1},M_{P_2})$ as a function of $M_{P_1}$ and the mass difference 
$\Delta M = M_{P_2}-M_{P_1}$ for $\sin 2\delta = 0.5$ and $v_T = 4$~GeV.
\label{fig:FMM}}

\EPSFIGURE{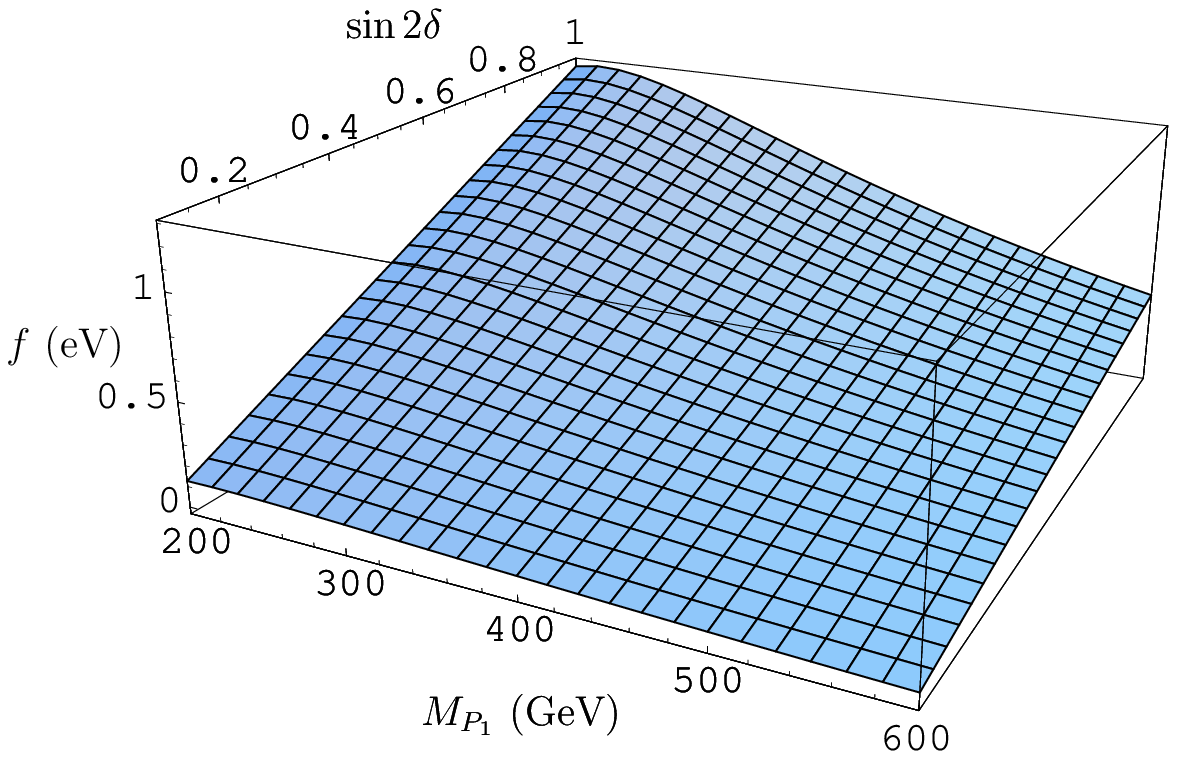,width=3.6in}
{The scale $f(M_{P_1},M_{P_2})$ as a function of $M_{P_1}$ and $\sin{2\delta}$ for 
$M_{P_2} = 1$~TeV and $v_T = 4$~GeV.
\label{fig:mixing}}

We plot the behavior of $f(M_{P_1},M_{P_2})$ as a function of $M_{P_1}$ and 
$\Delta M = M_{P_1}-M_{P_2}$ in Fig.~\ref{fig:FMM}. The range of parameters used is 
applicable to Case A. We see that overall, the neutrino mass increases as the mass 
difference of the two doubly charged scalar increases. In Fig.~\ref{fig:mixing} we 
plot the behavior of $f(M_{P_1},M_{P_2})$ for parameter range applicable to Case B. 
In both cases we expect neutrino masses to be in the sub-eV range.

We proceed next to examine how the neutrino oscillation data can constrains our 
model. Neutrino oscillations depend on the difference of mass-squared, hence we 
will focus on $m_{\nu}^2$ accordingly. Since the eigenvalues of $m_{\nu}^2$ are 
in general complex, and it is customary to separate out a phase matrix, we write
\begin{equation}\label{nudia}
m_\nu^2 = V^{T}_{PMNS}\,U
\begin{pmatrix}
m_1^{2} & 0 & 0 \\
0 & m_2^2 & 0 \\
0 & 0 & m_3^2
\end{pmatrix}
U\,V_{PMNS} \,, \qquad
U =
\begin{pmatrix} 
1 & 0 & 0 \\
0 & e^{i\varphi_1} & 0 \\
0 & 0 & e^{i(\varphi_2+\alpha)} 
\end{pmatrix} \,,
\end{equation}
where $V_{PMNS}$ is the neutrino mixing matrix~\cite{PMNS} and in standard notation
is the same as the quark mixing matrix \cite{PDG}, 
whereas  $\alpha$ is the Dirac phase. For normal hierarchy, we have $m_1 \simeq 0$, 
$m_2^2\simeq\Delta m^2_{\odot}$, and $m_3^2\simeq\Delta m_{atm}^2$. Oscillation 
experiments currently place limits on the following relevant 
parameters~\cite{nurev}: 
\begin{gather}
7.1 \times 10^{-5} < \Delta m^2_\odot < 8.9 \times 10^{-5}\,(\mathrm{eV}^2) 
\,, \qquad 0.164 < \sin^2{\theta_{12}} < 0.494 \,, \notag \\
1.4 \times 10^{-3} < |\Delta m_{atm}^2| < 3.3 \times10^{-3}\,(\mathrm{eV}^2) 
\,, \qquad 0.22 < \sin^2{\theta_{23}} < 0.85 \,, \notag \\
\sin^2{2\theta_{13}} = 0 \pm 0.04 \,.
\end{gather}
Using Eqs.~(\ref{numatrix}) and~(\ref{nudia}) and the oscillation data we can get
six constraints on the elements of $m^2_{\nu}$. From the first row of $m_{\nu}^2$ 
we obtain
\beq\label{etbd}
f^2\,Y_{e\tau}^2 \leq 1.32 \times 10^{2}\,\mathrm{eV}^{2} \,, \qquad
f^2\,Y_{e\tau}Y_{\mu\tau} \leq 1\,\mathrm{eV}^{2} \,, \qquad
f^2\,Y_{e\tau}Y_{\tau\tau} \leq 9.0 \times 10^{-2}\,\mathrm{eV}^{2} \,.
\end{equation}
Since the phases involved are unknown, we do not get lower bounds for these 
quantities. As will be seen below, the bounds given in 
Eq.~\eqref{etbd} are very loose compared to that found from rare muon and $\tau$ 
decays. The remaining three constraints are more stringent relatively, and they 
came by demanding a good fit to normal hierarchy:
\begin{align}
f^2(Y_{\mu\mu}^2 + 300 Y_{\mu\tau}^2) &\leq 2.7\,\mathrm{eV}^{2} \,, \\
\label{cyut}
f^2(Y_{\mu\mu} + 285Y_{\tau\tau})Y_{\mu\tau} &\leq 2.4 \times 10^{-1}
\,\mathrm{eV}^{2} \,, \\
\label{cytt}
f^2(Y_{\mu\tau}^2 + 278Y_{\tau\tau}^2) &\leq 2.9 \times 10^{-2}
\,\mathrm{eV}^{2} \,.
\end{align}
Using Eqs.~\eqref{cyut} and~\eqref{cytt}, we plot in Fig.~\ref{fig:utosci} the 
allowed parameter space for $Y_{\mu\tau}$ and $Y_{\tau\tau}$, with $f$ set to 
$0.5$~eV.~\footnote{In plotting Fig.~\ref{fig:utosci}, we have used
the fact that $Y_{\mu\mu}Y_{\mu\tau} \ll 0.24$, which we show below.}

\EPSFIGURE{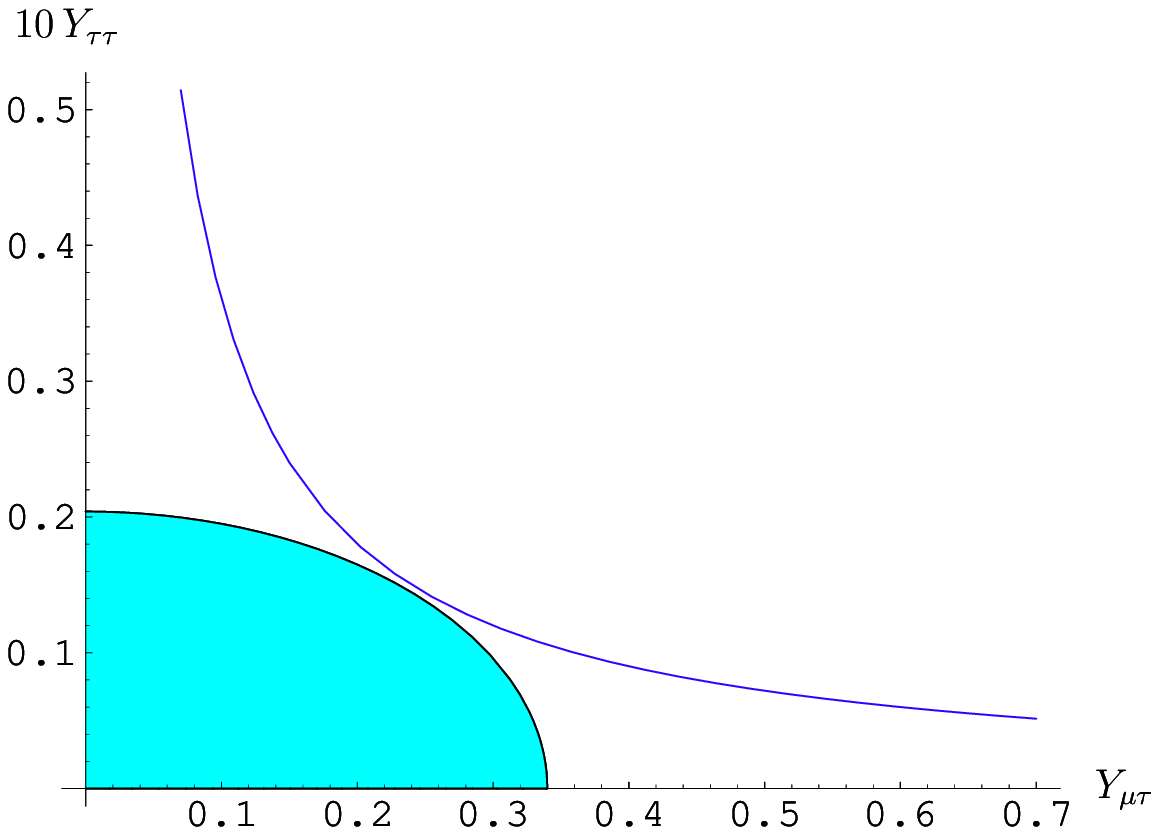,width=3.6in}
{Parameter space allowed for $Y_{\tau\tau}$ and $Y_{\mu\tau}$ for $f = 0.5$~eV. The 
hyperbolic curve is the upper limit set by Eq.~\eqref{cyut}. The shaded region is 
that allowed by both Eq.~\eqref{cyut} and Eq.~\eqref{cytt}.
\label{fig:utosci}}

\subsection{Rare muon and $\tau$ decays} 
The doubly charged Higgs bosons lead to many lepton number violating processes. 
Since no such signals were found in current experiments, they lead to very strong 
constraints on the Yukawa couplings. In the following, we work out these 
constraints.
\benu
\item {\bf Muonium anti-muonium conversion}

The effective Hamiltonian is given by the $P_{1,2}^{--}$ exchange at tree level:
\begin{equation}
H_{M\bar{M}}= \frac{Y_{ee}Y_{\mu\mu}}{2M_{--}^2}\,
\bar{\mu}\gamma^{\mu}e_R\,\bar{\mu}\gamma_\mu e_R + h.c. \,,
\end{equation}
where $M_{--}$ is the reduced mass of the pair of doubly charged Higgs given by
\beq
\frac{1}{M_{--}^{2}} = 
\frac{\sin^2 {\delta}}{M_{P_1}^{2}} + \frac{\cos^2 {\delta}}{M_{P_2}^{2}} \,.
\end{equation}
The current experimental limit~\cite{PDG} gives 
\beq
Y_{ee}Y_{\mu\mu} < 2.0 \times 10^{-3}\,(\Mred)^2 \,.
\end{equation}

\item {\bf Effective $e^{+} e^{-} \ra l^+l^-$, $l=e,\mu,\tau$, contact interactions}

The effective Hamiltonian for Bhabha scattering is
\begin{equation}
\frac{Y_{ee}^2}{M_{--}^2}\,\bar{e}_R\gamma^{\mu}e_R\,\bar{e}_R\gamma_{\mu}e_R \,.
\end{equation}
The bounds are 
\begin{align}\label{cti}
Y_{ee}^2 &< 1.8 \times 10^{-3}\,(\Mred)^2 \,, \notag \\
Y_{e\mu}^2 &< 2.4 \times 10^{-3}\,(\Mred)^2 \,, \notag \\
Y_{e\tau}^2 &< 2.4 \times 10^{-3}\,(\Mred)^2 \,.
\end{align}

\item {\bf Rare $\mu\ra 3e$ decays and its $\tau$ counterparts}

These decays can all be induced at the tree level and thus provide the most 
stringent limits on the Yukawa couplings. For $\mu \ra 3e$, the branching ratio 
is given by
\begin{equation}
Br(\mu \ra 3e)= 
\left(\frac{Y_{e\mu}Y_{ee}}{g^2}\right)^2 \left(\frac{M_W}{M_{--}}\right)^4 \,,
\end{equation}
with similar equations for $\tau$ decays. The constraints impose by the data is 
given by
\begin{align}\label{mu3e}
Y_{e\mu}Y_{ee} &< 6.6 \times 10^{-7}\,(\Mred)^2 \,, \notag \\
Y_{e\tau}Y_{ee} &< 3.0 \times 10^{-4}\,(\Mred)^2 \,, \notag \\
Y_{e\tau}Y_{\mu\mu} &< 3.0 \times 10^{-4}\,(\Mred)^2 \,, \notag \\
Y_{\mu\tau}Y_{\mu\mu} &< 2.9 \times 10^{-4}\,(\Mred)^2 \,, \notag \\
Y_{\mu\tau}Y_{ee} &< 2.9 \times 10^{-4}\,(\Mred)^2 \,.
\end{align}

\item {\bf Radiative flavor violating charged leptonic decays}

We consider here rare radiative decays of $\mu \ra e\gamma$ and 
$\tau \ra \mu (e) \gamma$. The fact that they are not seen to a very high precision
make them of paramount importance for probing the physics of lepton flavor 
violation. The branching for $\mu \ra e \gamma$ is calculated in~\cite{rare}
using an effective theory approach: 
\begin{equation}
Br(\mu \ra e \gamma) =
\frac{\alpha}{3\pi G_F^2}\sum_{l = e,\mu,\tau}
\left(\frac{Y_{l\mu}Y_{le}}{M_{--}^2}\right)^2 
\end{equation}
with obvious substitutions for $\tau$ decays.
The limits are given by
\begin{align}
\sum_l Y_{l\mu}Y_{le} &< 1.5 \times 10^{-5}\,(\Mred)^2 \,, \notag \\
\sum_l Y_{l\tau}Y_{le} &< 1.4 \times 10^{-3}\,(\Mred)^2 \,, \notag \\
\sum_l Y_{l\tau}Y_{l\mu} &< 1.1 \times 10^{-3}\,(\Mred)^2 \,.
\end{align}
\eenu

Comparing the sets of constraints we find that Eq.~(\ref{mu3e}) gives the strongest
limits. Although the limits from contact interactions, Eq.~({\ref{cti}), are less 
stringent, they are useful nonetheless as they constrain individual couplings. We
illustrate in Fig.~\ref{fig:etee} how $Y_{e\tau}$ and $Y_{ee}$ are restricted by 
both contact and rare decay experiments for a chosen value of the reduced mass 
$M_{--}=400$ GeV.

\EPSFIGURE{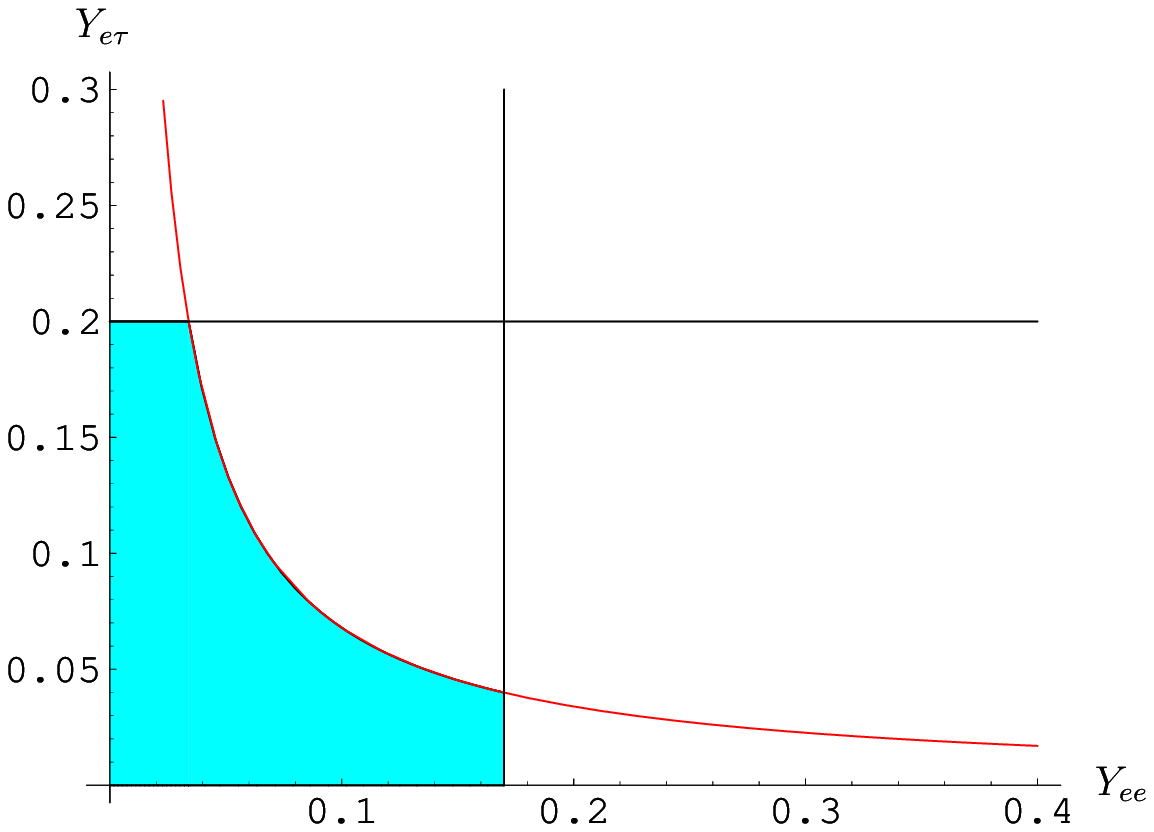,width=3.6in}
{Parameter space allowed for $Y_{e\tau}$ and $Y_{ee}$ (shaded). The straight lines 
are limits from the contact interactions, the hyperbolic curve the upper limit 
from $\tau\ra 3e$ decays. 
\label{fig:etee}}

It is also interesting to compare the limits from the rare decays with that from
neutrino oscillations. We plot in Fig.~\ref{fig:utuu} the parameter space allowed
in this case for $Y_{\mu\tau}$ and $Y_{\mu\mu}$, with $f = 0.5$~eV and 
$M_{--}=400$~GeV. Other comparisons are less instructive.

\EPSFIGURE{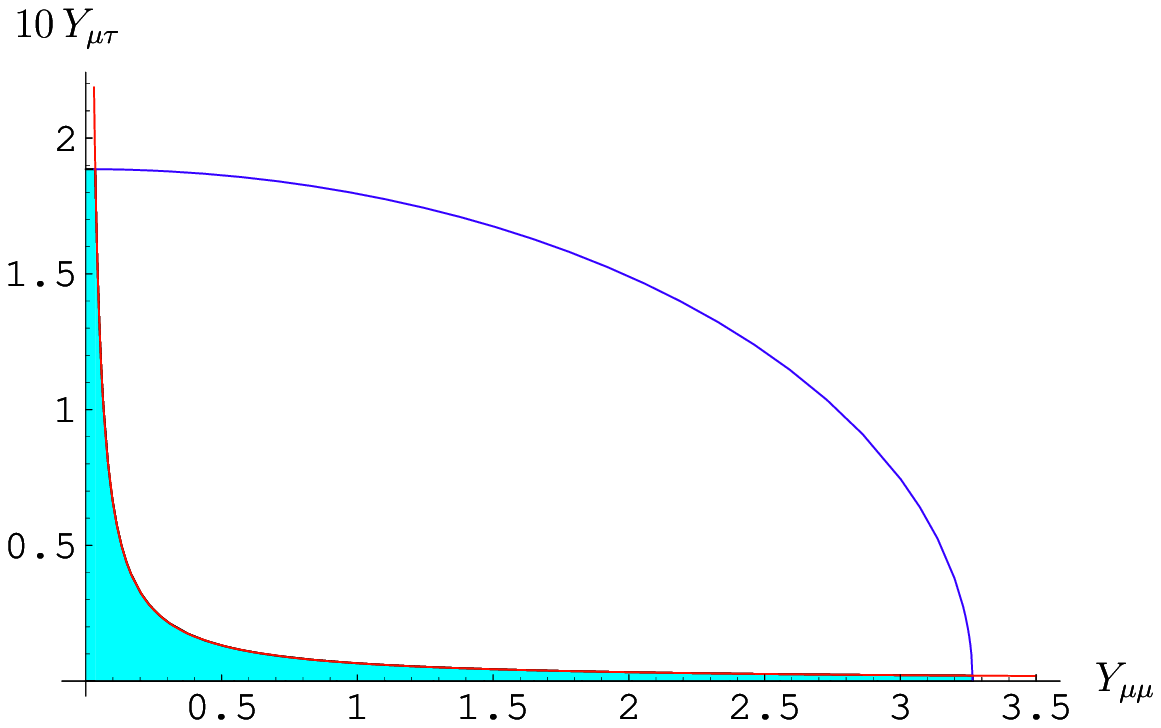,width=3.6in}
{Parameter space allowed for for $Y_{\mu\tau}$ and $Y_{\mu\mu}$ (shaded). The  
hyperbolic curve is the upper limit from $\tau \ra 3\mu$ decays, the ellipse the 
limit from fitting neutrino data.
\label{fig:utuu}}

In summary, we have from contact interactions the upper limits
\begin{equation}
\label{ubYe}
Y_{ee} < 0.17 \,, \qquad Y_{e\mu} < 0.2 \,, \qquad Y_{e\tau} < 0.2 \,, 
\end{equation}
and from neutrino data (see Fig.~\ref{fig:utosci} and~\ref{fig:utuu})
\begin {equation}
\label{ubYm}
Y_{\mu\mu} < 3.5 \,, \qquad Y_{\mu\tau} < 0.2 \,, \qquad Y_{\tau\tau}<0.02 \,.
\end{equation}
The values $M_{--}=400$~GeV and $f = 0.5$~eV are used throughout in obtaining these 
limits. They are consistent with and typical of what can be expected for our model.

\subsection{$\nonu$ decays of nuclei}

In our model, $\nonu$ decays of nuclei are induced by the exchanges of virtual 
$P_{1,2}^{--}$ bosons and Majorana neutrino as depicted in Fig.~\ref{fig:0nubb}.
\EPSFIGURE{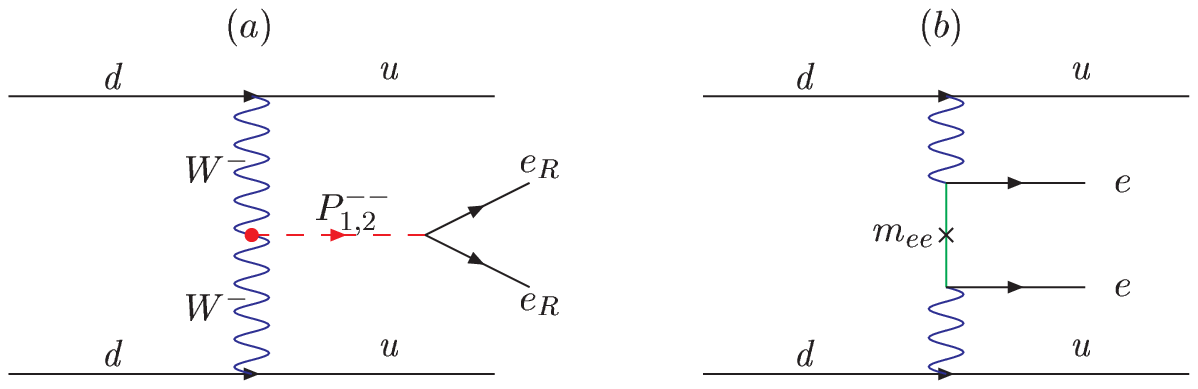,width=4in}
{$\nonu$ decays via exchange of: (a) doubly charged Higgs and (b) light Majorana 
neutrinos.
\label{fig:0nubb}} 
The quark level amplitude due to neutrino exchanges is given by~\footnote{In our
order of magnitude estimation, we have ignored spinor and kinematic factors, as 
well as factors from nuclear physics.}
\begin{equation}
A_{\nu}\sim\frac{g^4}{M_W^4}\frac{m_{ee}}{<p>^2} \,,
\end{equation}
where $<p>$ is the average momentum of the light neutrino exchanged. For notational
simplicity we will drop the subscript $\nu$ for the mass matrix. Typically,
$<p> \sim 0.1$~GeV which reflects the long range nature of light particle 
exchanges. Note that there is a cancellation between the contributions from 
$P_1$ and $P_2$, which is characteristic of a two level system.

The doubly charged Higgs exchange amplitude is given by
\begin{equation}
A_{P_{1,2}^{--}}\sim\frac{g^4\,Y_{ee} v_T \sin{2\delta}}{16\sqrt{2}M_W^4}
\left(\frac{1}{M_{P_1}^2}-\frac{1}{M_{P_2}^2}\right) \,,
\end{equation}
where $m_{ee}$ is given by Eq.~(\ref{numatrix}). We estimate that 
$A_{\nu}/A_{P_{1,2}^{--}} \lesssim 10^{-7}$. The smallness of this ratio is
due to the fact that in our model, $m_{ee}$ is suppressed not only by a two-loop 
factor, it is also suppressed by the electron mass factor $(m_e/M_W)^2$ coming 
from the doubly charged scalar coupling. We conclude that if seen, $\nonu$ decays 
of nuclei will be due to the existence of doubly charged Higgs at the weak scale. 
This can be tested at the LHC, and is the subject of next section. Since there is 
no conclusive evidence for these decays, we use it to set a limit of $Y_{ee}$. The
result is displayed in Fig.~\ref{fig:0vbb}. It can be seen that the limit here is 
comparable to that from the contact interactions.
\EPSFIGURE{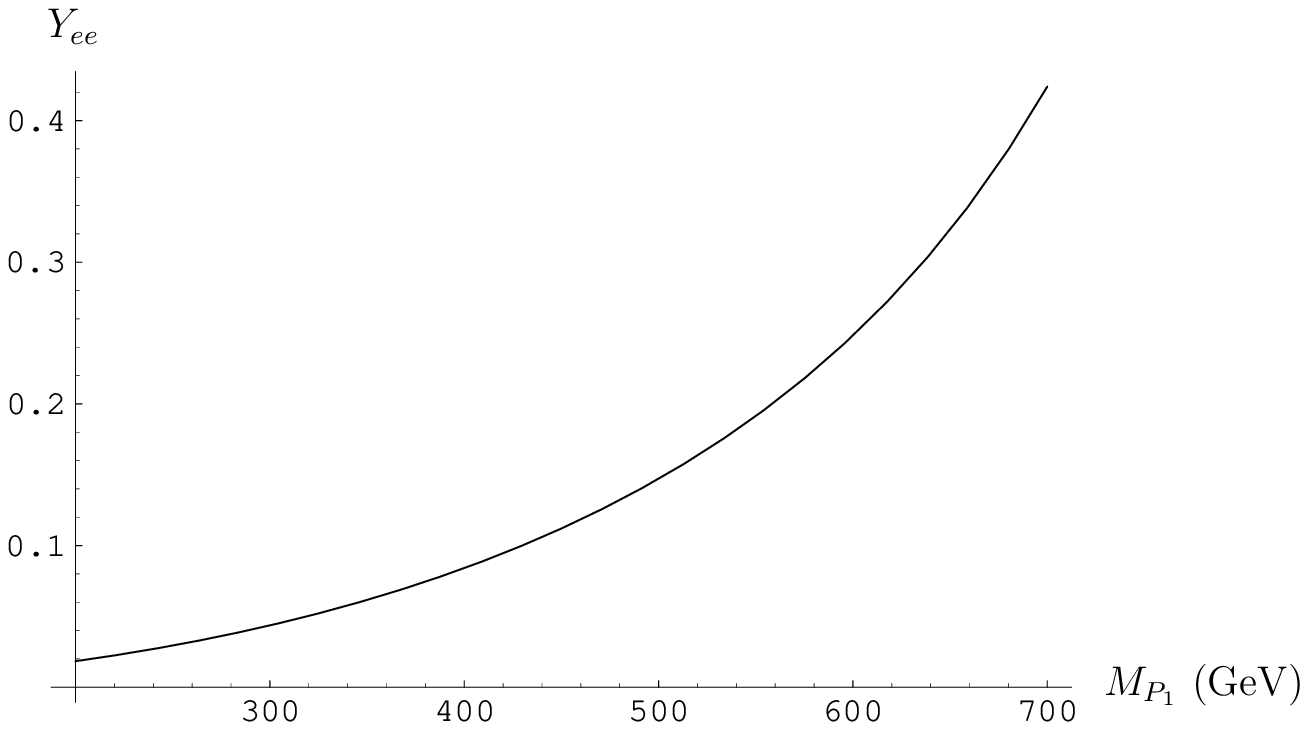,width=4in}
{Upper limit on $Y_{ee}$ as a function of $M_{P_1}$ for $\sin{2\delta} = 0.5$ and
$M_{P_2} = 1$~TeV.
\label{fig:0vbb}}

\section{Doubly Charged Higgs at the LHC} 
\subsection{Production of the doubly charged Higgs}\label{sec:LHCprod}
A central ingredient in our neutrino mass generation is the two doubly charged 
Higgs, $P_{1,2}^{\pm\pm}$. We have argued that if not both, at least one of the
doubly charged Higgs is well within reach of the LHC. Without loss of generality, 
we will take $P_1$ to be this (lighter) state, and focus on its production and 
decay below.

Now the doubly charged Higgs, $P_{1,2}^{\pm\pm}$, have no direct couplings to the
quarks, which is characteristic of models with doubly charged scalars. However, 
they do couple to the SM gauge bosons. Thus at the LHC, $P_1^{\pm\pm}$ will be 
produced predominantly via the $WW$ fusion processes, as illustrated in 
Fig.~\ref{fig:0nubb}(a) (the d-quark is replaced by the u-quark for $P_1^{++}$), 
and the Drell-Yan (DY) annihilation processes,
 \begin{equation}
q\bar{q} \ra \gamma^*,Z^*\ra P_1^{++}P_1^{--} \quad (q = u,d) \,.
\end{equation}
 The 
relevant gauge-scalar couplings are given by~\footnote{For couplings with $P_2$, 
$c_\delta \ra -s_\delta$, $s_\delta \ra c_\delta$.}
\begin{align}\label{prodcoups}
W_{\mu}^{\pm}W_{\nu}^{\pm}P_1^{\mp\mp} &:
\frac{g^2}{\sqrt{2}}\,v_T\,c_\delta\,W_{\mu}^{+}W_{\nu}^{+}P_1^{--} + h.c. \notag \\
A_{\mu}\,P_1^{++}P_1^{--} &: 
i\,2e\,A_{\mu}\,\PD_{\nu}{P_1^{++}}P_1^{--} + h.c. \notag \\
Z_{\mu}\,P_1^{++}P_1^{--} &:
\frac{ig}{c_W}\!\left[(1-2s_W^2)c_\delta^2-2s_W^2 s_\delta^2\right]\!
Z_{\mu}\,\PD_{\nu}{P_1^{++}}P_1^{--} + h.c. 
\end{align}
where $g = e/\sin\theta_W$, $s_W\equiv\sin\theta_W$, $c_\delta\equiv\cos\delta$ and 
$s_\delta\equiv\sin\delta$.

We calculate the production cross sections numerically using the CalcHEP 
package~\cite{CalcHEP}, which allows an easy implementation of our model. The 
calculations are leading order, and are done in unitary gauge. The cross sections
are calculated for pp collisions in the center-of-mass (CM) frame with energy 
$\sqrt{s} = 14$~TeV, and CTEQ6M~\cite{Pumplin} parton distributions functions are 
used to fold in the cross section for the hard partonic production processes. 

 From Eq.~\eqref{prodcoups} we see that the only model parameters the production 
cross section explicitly depend on are $v_T$ and the mixing angle $\delta$. The
dependence on all other model parameters are implicit through the dependence on the
$P_1^{\pm\pm}$ mass, $M_{P_1}$, as given in Eq.~\eqref{DCH}. The choice we made in
our calculations is thus to set 
\beq\label{pchoice}
M = v_T \,, \quad \kappa_\Psi = -\kappa_2 \,, \quad 
m = 2 v = 492.442\,\mathrm{GeV} \,, \quad \lambda = -\lambda_T' = \rho/2 = 1 \,, 
\eeq
and vary $M_{P_1}$ by varying $\kappa_2$. Note that with this choice, $s_\delta$ is
independent of $\kappa_2$, and thus $M_{P_1}$. For the SM parameters, we take 
$v = 246.221$~GeV, $e = 0.3133$ and $s_W = 0.4723$.

\EPSFIGURE{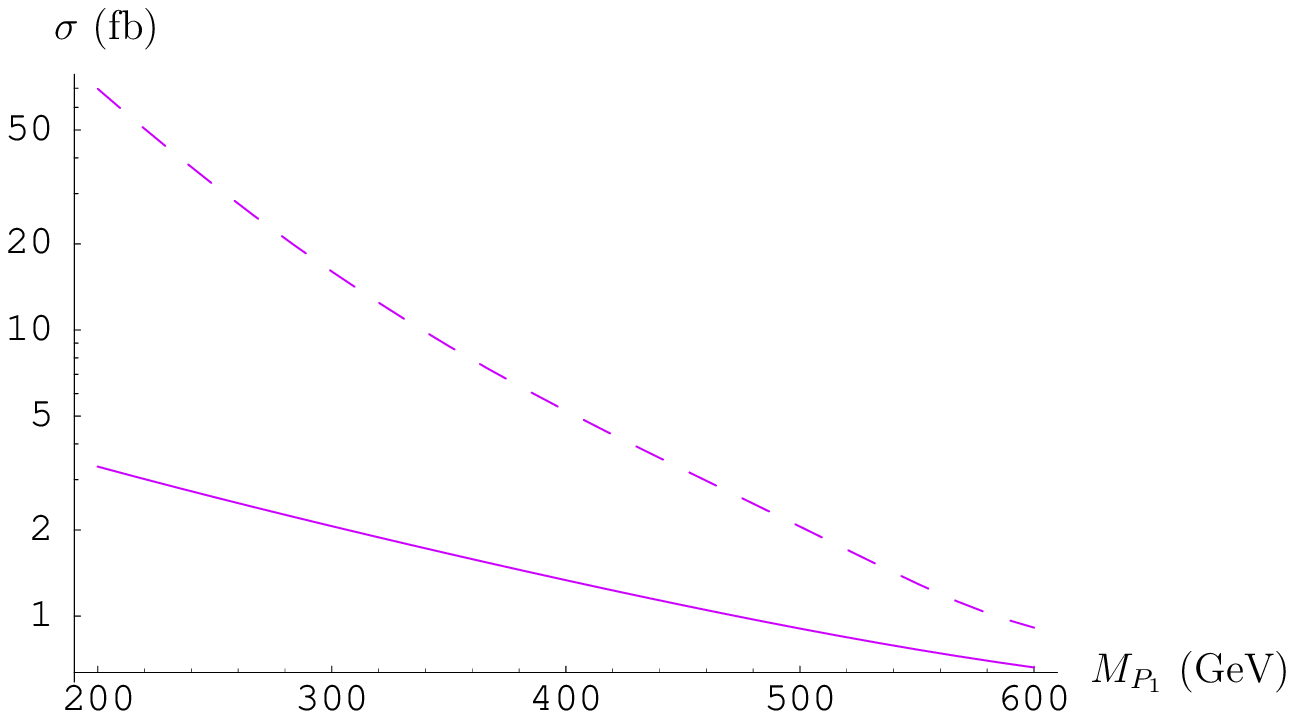,width=4in}
{Production cross sections for $P_1^{\pm\pm}$ with $v_T = 4$~GeV and 
$s_\delta = 0.12$. The solid line traces the results from the W fusion 
processes, the dashed line the Drell-Yan annihilation processes.
\label{fig:prodxs}}

We plot in Fig.~\ref{fig:prodxs} the production cross section for $P_1^{\pm\pm}$ 
from $WW$ fusion and Drell-Yan pair annihilation as a function of $M_{P_1}$. Note 
that QCD corrections are expected to increase the DY production cross section 
by a factor of about 1.25 at next-to-leading order~\cite{DYQCD}. The CalcHEP results
are checked with that calculated from \textsc{Pythia}~\cite{pythia}, and are found 
to be consistent. 

Except for the mixing angle factors $c_\delta$ and $s_\delta$, the coupling of 
$P_1^{\pm\pm}$ to the weak gauge bosons are the same as that of $\Delta_L^{\pm\pm}$ 
to the $W_L$, $Z_1$ bosons in the left-right symmetric model~\cite{Gunion}, while 
the couplings to photon is the same in both models. Thus, the results for the cross 
section should agree once the scaling factors are taken into account. Comparing the
results of our model to that of Refs.~\cite{Huitu,Azuelos}, we find good agreements.

We see from Fig.~\ref{fig:prodxs} that DY pair production is dominant compared to 
the single production channel via $WW$ fusion for the whole range of $P_1^{\pm\pm}$ 
masses in our model. This is a robust feature in the production of the doubly 
charged Higgs at hadron colliders in many models. This is because the virtual
photon exchange is the same for all models and only the $Z$ exchange term is model 
dependent, which is subdominant. The relative magnitude of production via WW 
fusion compared to the DY pair production can thus be used to distinguish between
the different models containing the doubly charged Higgs.

\subsection{The decay of $P_1^{\pm\pm}$}

At the leading order, there are six decay channels possible for $P_1^{\pm\pm}$:
\begin{align}\label{decay}
(1)\;P_1^{\pm\pm} &\ra l_{aR}^\pm l_{bR}^{\pm} \quad (a,b = e,\mu,\tau)\,,\notag \\
(2)\;P_1^{\pm\pm} &\ra W^{\pm}W^{\pm}\,,\notag \\
(3)\;P_1^{\pm\pm} &\ra P^{\pm}W^{\pm}\,,\notag \\
(4)\;P_1^{\pm\pm} &\ra P^{\pm}P^{\pm}\,,\notag \\
(5)\;P_1^{\pm\pm} &\ra W^{\pm}W^{\pm}X^0\,, \qquad X^0 = T^0_a,h^0,P^0 \notag \\
(6)\;P_1^{\pm\pm} &\ra P^{\pm}P^{\pm}X^0\,.
\end{align}
Kinematically, mode (4) and (6) are not allowed in our model, while the 
availability of the rest depends on the value of the scalar boson masses. The 
coupling for mode (2) has been given in Eq.~\eqref{prodcoups}, for mode (1), (3) 
and (5) the couplings are given by
\begin{align}\label{decaycoups}
P_1^{\pm\pm}l_{aR}^\mp l_{bR}^\mp &: 
Y_{ab}\,s_\delta P_1^{--}\,\overline{l_{aR}^c}\,l_{bR} + h.c. \notag \\
P_1^{\pm\pm}W_{\mu}^{\mp}P^{\mp} &: 
i g\,c_\delta\,W_{\mu}^-\!\left[\PD_{\nu}P_1^{++}P^{-}-P_1^{++}\PD_{\nu}P^{-}\right]
+ h.c. \notag \\
P_1^{\pm\pm}W_{\mu}^{\mp}W_{\nu}^{\mp}X^0 &:
\frac{g^2}{\sqrt{2}}\,c_\delta\,c_X P_1^{\pm\pm}W_{\mu}^{\mp}W_{\nu}^{\mp}X^0
+ h.c.
\end{align}
where $c_X = i,\,\cos\vartheta,\,\sin\vartheta$ for $X^0 = T^0_a,h^0,P^0$ 
respectively. 

The leptonic decays are kinematically the most favorable mode of decay for the 
doubly charge Higgs, $P_1^{\pm\pm}$, and this is a universal feature in any model 
that contains them. The leptonic decay width has a very simple form given by
\begin{equation}\label{lepDW}
\Gamma(l_{aR}^\pm l_{bR}^\pm) = 
(1 + \delta_{ab})\frac{|Y_{ab}|^2}{16\pi}s_\delta^2\,M_{P_1} \quad
(\textrm{no sum}) \,.
\end{equation}
Note that in our model, the final state charged leptons are right-handed. Hence in 
principle, helicity measurements can be used to distinguish between our model and 
those whose doubly charged Higgs coupling only to left-handed leptons (see 
e.g.~\cite{LLH,AA05,AAO06,Hektor,Han}).

 From the discussion above, it is not unreasonable to take $Y_{ab} \sim 0.1$ (except 
for $Y_{\tau\tau}$, which is constrained to be ten times smaller). Thus, provided
the mixing angle between the doubly charged Higgs bosons, $\delta$, is not too 
small, one can expect spectacular signals from like-sign dileptons such 
as $e\mu$, $e\tau$ and $\mu\tau$, that are directly produced from $P_1^{\pm\pm}$.

The $W^{\pm}W^{\pm}$ channel opens up once $M_{P_1} > 2M_W$. Its decay width is 
\beq
\Gamma(W^\pm W^\pm)= 
\frac{g^4 v_T^2 c_\delta^2}{16\pi M_{P_1}}\sqrt{1-\frac{4M_W^2}{M_{P_1}^2}}
\left(3-\frac{M_{P_1}^2}{M_W^2}+\frac{M_{P_1}^4}{4M_W^4}\right) \,,
\eeq
which is proportional to $v_T^2$. Given that $v_T$ is small, and $M_{P_1}$ is of 
order the electroweak scale, whether the leptonic or $W^{\pm}W^{\pm}$ mode dominates
depends on the value of the mixing angle $\delta$.

\EPSFIGURE{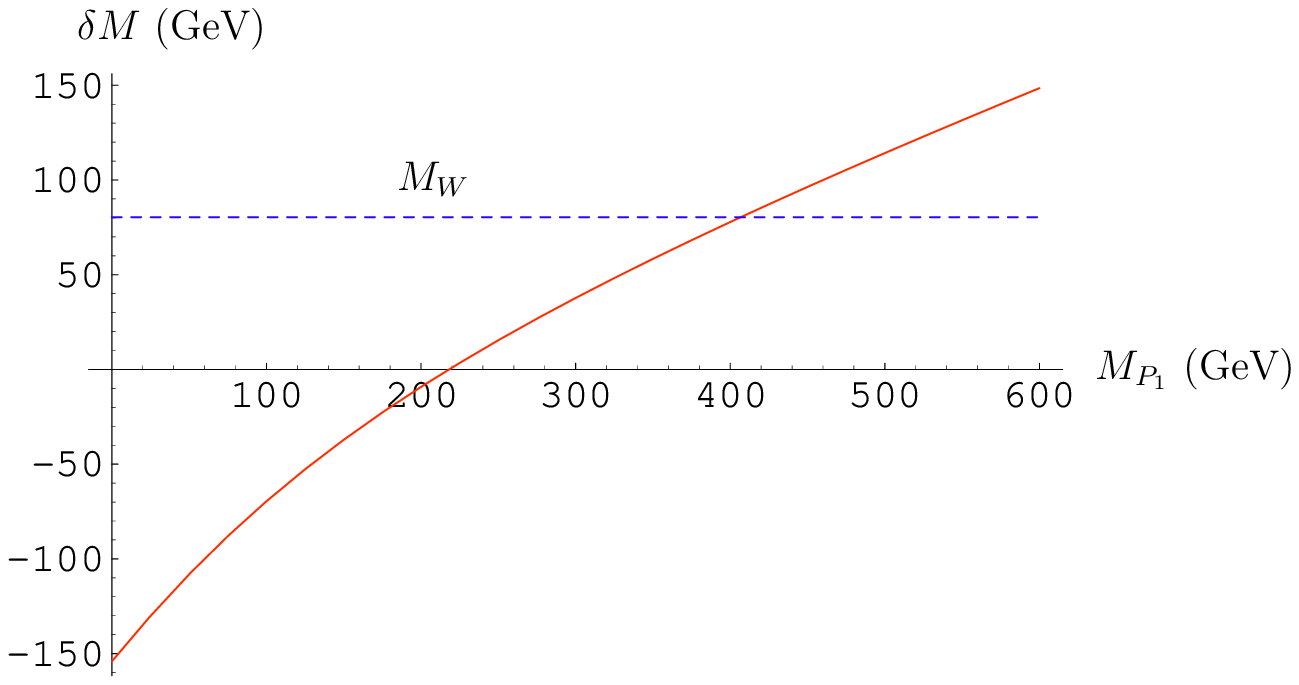,width=4in}
{The mass difference $\delta M = M_{P_1}-M_P$ as a function of $M_{P_1}$ for 
$v_T = 4$~GeV. The short dashed line marks where $\delta M = M_W$, which opens 
up the decay channel.
\label{fig:dM}}

The $W^{\pm}P^{\pm}$ mode opens up once the mass difference, 
$\delta M = M_{P_1}-M_P$, where $M_P$ is the mass of the singly charged Higgs boson,
is greater than $2M_W$. From Eq.~\eqref{MassP}, we see that if $M_P$ is to be of 
interest at the LHC, $\omega$ needs to be order unity (the validity of perturbation
theory constrains $|\kappa_2|$ to be less than $4\pi$). This implies that 
$M \sim v_T$, and in Fig.~\ref{fig:dM} we plot $\delta M$ as a function of $M_{P_1}$
for $M = v_T =4$~GeV. We vary $M_{P_1}$ the same way as detailed in 
Sec.~\ref{sec:LHCprod}, with the same choice of parameters, Eq.~\eqref{pchoice}.
For this choice, the decay channel opens up once $M_{P_1} > 407$~GeV. 

The decay width for the $W^{\pm}P^{\pm}$ mode is given by
\beq
\Gamma(W^\pm P^\pm) =
\frac{g^2 c_\delta^2 M_{P_1}^3}{16\pi M_W^2}\lambda^\frac{3}{2}\!
\left(1,\frac{M_W^2}{M_{P_1}^2},\frac{M_P^2}{M_{P_1}^2}\right) \,,
\eeq
where $\lambda(x,y,z) = x^2+y^2+z^2-2xy-2xz-2yz$. With a dependence of $g^2$ and no
suppression coming from factors of $v_T$, the $W^{\pm}P^{\pm}$ mode is expected to
dominate over the $W^{\pm}W^{\pm}$ mode once it becomes available. Note that the 
singly charged Higgs decays primarily into $W\gamma$ rather than fermion pairs, 
which is the dominant decay mode for the usual charged Higgs originating from SM 
doublets.

In general, the three-body decay mode $W^{\pm}W^{\pm}X^0$, $X^0 = T^0_a,h^0,P^0$,
is expected to be relatively suppressed by the three-body phase space when compared
to the two-body modes. However, since the couplings do not depend on $v_T$, if 
$s_\delta \ll 1$, it may become important compared to the lepton and 
$W^{\pm}W^{\pm}$ modes. 

The masses of the neutral pseudoscalar and scalar bosons, $m_{X^0}$, are given
by Eqs.~\eqref{psmass} and~\eqref{smass}. Here too, $\omega\sim\mathcal{O}(1)$ is 
required to keep the neutral bosons at electroweak scale. Note that once $\omega$ 
is fixed, $M_{X^0}$ and $M_{P_1}$ are governed by independent sets of parameters.
Below, we will choose $X^0 = T^0_a$ as the representative three-body mode, which
has the least and simplest model dependence compared to the neutral scalar cases. 
The difference in the decay widths would come in only as ratios of masses, and from 
the neutral mixing angle factors, $c_\vartheta$ and $s_\vartheta$. 


\EPSFIGURE{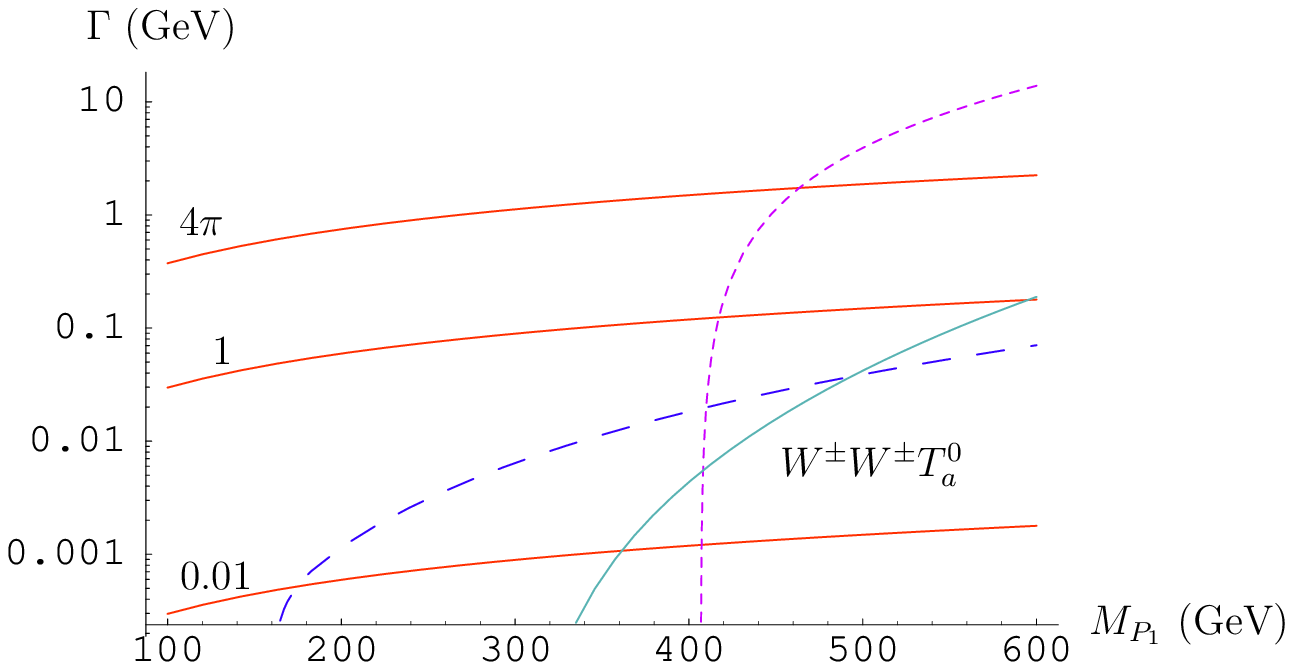,width=4.2in}
{Decay widths of $P_1^{\pm\pm}$ as a function of $M_{P_1}$, with $v_T= 4$~GeV and
the mixing angle fixed at $\sin\delta = 0.12$. The leptonic mode is represented by 
the solid lines at different values of $|Y_{ab}|^2$, the $W^{\pm}W^{\pm}$ mode the 
long dashed line, and the $W^{\pm}P^{\pm}$ mode the short dashed line. The 
three-body $W^{\pm}W^{\pm}T^0_a$ mode is labelled on the plot.
\label{fig:DWidth}}

For the choice $M = v_T = 4$~GeV, $M_{T^0_a} = 146.5$~GeV, and the  
$W^{\pm}W^{\pm}T^0_a$ mode opens up once $M_{P_1} > 307.31$~GeV. Since there is no 
known closed form for the phase space of general three-body decays, we calculate the 
three-body decay widths numerically using CalcHEP as a function of $M_{P_1}$. We 
display the result and compared with all the allowed two-body modes listed in 
Eq.~\eqref{decay} in Fig.~\ref{fig:DWidth}. We work in the small mixing scenario 
where $s_\delta = 0.12$, since for $s_\delta \sim 1$, the results are qualitatively 
that of the left-right symmetric models~\cite{Gunion}.

There are several noteworthy features. First, despite the phase space advantage, 
the suppression of the lepton mode by the $s_{\delta}^2$ factor is such that the 
$W^{\pm}W^{\pm}$ mode is never negligible for $|Y_{ab}|^2 \sim 1$, and dominating
for $|Y_{ab}|^2 \ll 1$ even for small values of $M_{P_1}$. Next as discussed above,
the $W^{\pm}P^{\pm}$ mode dominates over both the lepton and $W^{\pm}W^{\pm}$ 
modes, and does so almost at once after it becomes kinematically available. 
Moreover, even if one were to push $|Y_{ab}|^2$ to the perturbative limit of $4\pi$, 
the $W^{\pm}P^{\pm}$ mode would still become dominant soon above its threshold. 
Lastly, and perhaps most interestingly, the three-body 
$W^{\pm}W^{\pm}T^0_a$ mode dominates over the lepton mode when $|Y_{ab}|^2 \ll 1$ 
(suppression by $s_\delta$), and is significant relative to the $W^{\pm}W^{\pm}$
mode and even overtakes it at large values of $M_{P_1}$ (suppression by $v_T$). 

These are features that are not shared by other doubly charged Higgs models, and 
will serve to distinguish our model from them at the LHC.

\section{Conclusion}

We have given a detail study of the phenomenology of a model where Majorana neutrino
mass is induced radiatively at two-loop. This is made possible by extending the 
Higgs sector of the SM, and the new structure involves a triplet and a doubly 
charged singlet field. Lepton number is violated only in the scalar sector by a hard
and a soft term. When these two terms are set to zero, the model conserves lepton 
number, and is technically natural. 

Since the active neutrinos masses are two-loop effects due to new physics at the TeV
scale, the model is testable at the LHC. In our study, we have focused on the 
production and decays of the doubly charged Higgs bosons. We find it is possible in 
the enlarged Higgs sector parameter space to have always at least one doubly charged
Higgs with mass $\lesssim 600$~GeV. Furthermore, in much of the parameter space, two
doubly charge Higgs can be expected to have masses within reach of the LHC. We find
that Drell-Yan mechanism is the dominant production mechanism, with single 
production from $WW$ fusion coming into play only for masses greater than 600~GeV. 
This is common to most doubly charged Higgs models. We find cross sections ranging 
from $\mathcal{O}(1 - 70)$~fb without including QCD effects, which typically cause
an enhancement by a factor of 1.25. 

The decays of the doubly charged Higgs are potentially the best way of 
differentiating between the different doubly charged Higgs models. We find that with
reasonable values of the Yukawa couplings, the leptonic decays are not always 
dominant despite being kinematically favorable. Nevertheless, decay modes producing 
directly same sign dilepton pairs such as $e\tau$, $\mu\tau$, and $e\mu$, are 
distinctive and unmistakable. Moreover, since the doubly charged Higgs only couples 
to RH leptons in our model, their decays into a pair of tauons offer opportunities 
for helicity measurements. Another interesting feature of the doubly charged Higgs
decays is that three-body modes $W^{\pm}W^{\pm}X^0$ involving the neutral scalar and
pseudoscalar bosons can become as important as the two-body modes, particularly when
the mixing between the doubly charged Higgs is small.

The model we studied here can only accommodate the normal hierarchy for the active 
neutrino mass matrix. It predicts that the first element to be very small. Moreover,
$\nonu$ decays of nuclei can still occur via the exchange of doubly charged Higgs. 
This give rise to the interesting situation whereby if such decays are seen, the 
doubly charged Higgs will also be seen at the LHC, and vice versa. If the neutrino
mass hierarchy turns out to be other than the normal hierarchy, neutrino masses are 
not generated by the model we studied.

\acknowledgments{We thank Dr. D. Axen for kindly sharing with us his knowledge and 
insights on \textsc{Pythia}, and Dr. G. Azuelos for useful discussions.
This work is supported in part by the National Science Council of R.O.C. under 
Contract No. NSC-95-2112-M-007-059-MY3, and by the Natural Science and Engineering 
Council of Canada.}


\begin{thebibliography}{99}
\bibitem{HS}
K.~R.~Dienes, E.~Dudas and T.~Gherghetta,
\textit{Light neutrinos without heavy mass scales: A higher-dimensional seesaw mechanism},
\npb{557}{1999}{25} [\hepph{9811428}]; \\
S.~J.~Huber and Q.~Shafi,
\textit{Seesaw mechanism in warped geometry},
\plb{583}{2004}{293}. 

\bibitem{Zee}
A.~Zee,
\textit{A Theory Of Lepton Number Violation, Neutrino Majorana Mass, And Oscillation},
\plb{93}{1980}{389} [Erratum-ibid.\ {\bf B 95} (1980) 461];\\
\textit{Charged Scalar Field And Quantum Number Violations},
\plb{161}{1985}{141}.

\bibitem{nurev}
G.~L.~Fogli, E.~Lisi, A.~Marrone and A.~Palazzo,
\textit{Global analysis of three-flavor neutrino masses and mixings},
\ppnp{57}{2006}{742} [\hepph{0506083}].

\bibitem{Zee2}
A.~Zee, 
\textit{Quantum Numbers Of Majorana Neutrino Masses},
\npb{264}{1986}{99}; \\
K.S.~Babu,
\textit{Model of 'Calculable' Majorana Neutrino Masses},
\plb{203}{1988}{132}.

\bibitem{P1}
C.~S.~Chen, C.~Q.~Geng and J.~N.~Ng,
\textit{Unconventional neutrino mass generation, neutrinoless double beta decays, and collider phenomenology},
\prd{75}{2007}{053004} [\hepph{0610118}].

\bibitem{PDG}
W.~M.~Yao {\it et al.} [Particle Data Group],
\textit{Review of particle physics},
\jphg{33}{2006}{1}.

\bibitem{CDF}
O.~Stelzer-Chilton [for the CDF Collaboration],
\textit{First Run II Measurement of the W Boson Mass with CDF},
\hepex{0706.0284}.
  
\bibitem{GN}
M.~C.~Gonzalez-Garcia and Y.~Nir,
\textit{Implications Of A Precise Measurement Of The Z Width On The Spontaneous Breaking Of Global Symmetries},
\plb{232}{1989}{383}.

\bibitem{2loop}
J.~van der Bij and M.~J.~G.~Veltman,
\textit{Two Loop Large Higgs Mass Correction To The Rho Parameter},
\npb{231}{1984}{205}; \\
K.~L.~McDonald and B.~H.~J.~McKellar,
\textit{Evaluating the two loop diagram responsible for neutrino mass in Babu's model},
\hepph{0309270}.

\bibitem{2loopl}
K.~S.~Babu and C.~Macesanu,
\textit{Two-loop neutrino mass generation and its experimental consequences},
\prd{67}{2003}{073010} [\hepph{0212058}].

\bibitem{PMNS}
B.~Pontecorvo,
\textit{Neutrino experiments and the question of leptonic-charge  conservation},
\jetp{26}{1968}{984}; \\
Z.~Maki, M.~Nakagawa, and S.~Sakata, 
\textit{Remarks on the unified model of elementary particles},
\ptp{28}{1962}{870}.

\bibitem{rare}
M.~Raidal and A.~Santamaria, 
\textit{$\mu$--$e$ conversion in nuclei versus $\mu \ra e\gamma$: An effective field theory point of view},
\plb{421}{1998}{250}

\bibitem{CalcHEP}
A.~Pukhov {\it et al.},
\textit{CompHEP: A package for evaluation of Feynman diagrams and integration over multi-particle phase space. 
User's manual for version 33},
\hepph{9908288};
A.~Pukhov,
\textit{CalcHEP 3.2: MSSM, structure functions, event generation, batchs, and generation of matrix elements for 
other packages},
\hepph{0412191}.

\bibitem{Pumplin}
J.~Pumplin, D.~R.~Stump, J.~Huston, H.~L.~Lai, P.~Nadolsky and W.~K.~Tung,
\textit{New generation of parton distributions with uncertainties from global QCD analysis},
\jhep{0207}{2002}{012} [\hepph{0201195}].

\bibitem{DYQCD}
M.~M\"uhleitner and M.~Spira,
\textit{Note on doubly charged Higgs boson pair production at hadron colliders},
\prd{68}{2003}{117701}

\bibitem{pythia}
T.~Sjostrand, S.~Mrenna and P.~Skands,
\textit{PYTHIA 6.4 physics and manual},
\jhep{0605}{2006}{026} [\hepph{0603175}].

\bibitem{Gunion}
J.~F.~Gunion, J.~Grifols, A.~Mendez, B.~Kayser and F.~I.~Olness,
\textit{Higgs Bosons In Left-Right Symmetric Models},
\prd{40}{1989}{1546}.

\bibitem{Huitu}
K.~Huitu, J.~Maalampi, A.~Pietila and M.~Raidal,
\textit{Doubly charged Higgs at LHC},
\npb{487}{1997}{27} [\hepph{9606311}].

\bibitem{Azuelos}
G.~Azuelos, K.~Benslama and J.~Ferland,
\textit{Prospects for the search for a doubly-charged Higgs in the left-right symmetric model with ATLAS},
\jphg{32}{2006}{73} [\hepph{0503096}].

\bibitem{LLH}
J.A.~Coarasa, A.~M\a' endez, and J.~Sol\a`a, 
\textit{Higgs triplet effects in purely leptonic processes}, 
\plb{374}{1996}{131}. 

\bibitem{AA05}
A.~G.~Akeroyd and M.~Aoki,
\textit{Single and pair production of doubly charged Higgs bosons at hadron colliders},
\prd{72}{2005}{035011} [\hepph{0506176}].

\bibitem{AAO06}
A.~G.~Akeroyd, M.~Aoki and Y.~Okada,
\textit{Lepton Flavour Violating tau Decays in the Left-Right Symmetric Model},
\hepph{0610344}.

\bibitem{Hektor}
A.~Hektor, M.~Kadastik, M.~Muntel, M.~Raidal and L.~Rebane,
\textit{Testing neutrino masses in little Higgs models via discovery of doubly charged Higgs at LHC},
\hepph{0705.1495}.

\bibitem{Han}
T.~Han, B.~Mukhopadhyaya, Z.~Si and K.~Wang,
\textit{Pair Production of Doubly-Charged Scalars: Neutrino Mass Constraints and Signals at the LHC},
\hepph{0706.0441}.
\end{thebibliography}
\end{document}